\documentclass[twocolumn,showpacs,floatfix]{revtex4}%
\usepackage{graphicx}
\usepackage{dcolumn}
\usepackage{bm}
\usepackage{amssymb}
\usepackage{amsmath}
\usepackage{amsfonts}
\usepackage{nameref}
\usepackage{float}
\begin{document}

\title{Electron localization in dissociating ${\textrm{H}_2}^+$ by retroaction of a photoelectron onto its source}

\author{M. Waitz$^{1}$} 
\author{D. Aslit\"urk$^{1}$}
\author{N. Wechselberger$^{1}$}
\author{H. K. Gill$^{1}$}
\author{J. Rist$^{1}$}
\author{F. Wiegandt$^{1}$}

\author{C. Goihl$^{1}$}
\author{G. Kastirke$^{1}$}
\author{M. Weller$^{1}$}
\author{T. Bauer$^{1}$}
\author{D. Metz$^{1}$}
\author{F. P. Sturm$^{1,3}$}
\author{J. Voigtsberger$^{1}$}
\author{S. Zeller$^{1}$}
\author{F. Trinter$^{1}$}

\author{G. Schiwietz$^{2}$}

\author{T. Weber$^{3}$}

\author{J. B. Williams$^{1}$}
\author{M. S. Sch\"offler$^{1}$}
\author{L. Ph. H. Schmidt$^{1}$}
\author{T. Jahnke$^{1}$}
\author{R. D\"orner$^{1}$}
\email{doerner@atom.uni-frankfurt.de}

\affiliation{
$^1$ Institut f\"ur Kernphysik, J.~W.~Goethe Universit\"at, Max-von-Laue-Str. 1, 60438 Frankfurt, Germany \\
$^2$ Helmholtz-Zentrum Berlin f\"ur Materialien und Energie, Institute G-ISRR, \mbox{Hahn-Meitner-Platz 1, 14109 Berlin, Germany} \\
$^3$ \mbox{Chemical Sciences Division, Lawrence Berkeley National Laboratory}, Berkeley, California 94720, USA \\
}

\date{\today}

\begin{abstract}

We investigate the dissociation of ${\textrm{H}_2}^+$ into a proton and a $\textrm{H}^0$ after single ionization with photons of an energy close to the threshold. 
We find that the $\textrm{p}^+$ and the $\textrm{H}^0$ do not emerge symmetrically in case of the ${\textrm{H}_2}^+$ dissociating  along the $1s\sigma_g$ ground state.
Instead, a preference for the ejection of the $\textrm{p}^+$ in the direction of the escaping photoelectron can be observed.
This symmetry breaking is strongest for very small electron energies. 
Our experiment is consistent with a recent prediction by Serov and Kheifets [Phys. Rev. A \textbf{89}, 031402 (2014)]. In their model, which treats the photoelectron classically, the symmetry breaking is induced by the retroaction of the long range Coulomb potential onto the dissociating ${\textrm{H}_2}^+$.

\end{abstract}

\maketitle

\section{Introduction}
 
Symmetry is one of the most fundamental concepts for the quantum mechanical description of molecules. Due to their symmetry, homonuclear molecules have electronic eigenfunctions of either gerade or ungerade parity. This is commonly assumed to hold true while a molecule is dissociating, even though after dissociation, any measurement will detect the hole at one of the fragments (i. e. a symmetry-broken system). The well defined symmetry of the electron wave function will, however, create a hole with equal probability at each of the two fragments.  
External fields which are present during the dissociation can break this symmetry. A variety of scenarios have been reported in which strong laser fields have been utilized to induce such symmetry breaking. A pioneering experiment used a carrier envelope phase locked few cycle pulse \cite{Kling06science}. Later experiments used two color pulses of neighboring harmonics \cite{Ray09prl,Wu13pra}, an attosecond pulse synchronized to the driving pulse \cite{Sansonenature2010,Singh10prl,Fischer10prl} or broke the laser field symmetry by measuring the field direction at the instant of creating an ${\textrm{H}_2}^+$ ion employing the attoclock technique \cite{Wu13natcomm}. In all these scenarios, the laser field mixed gerade and ungerade states with a well-defined phase during the dissociation leading to a localization of the bound electron. Such a coherent mixture of states of two symmetries can also occur already in the ionization step if the ionization energy is in the range of doubly excited resonances \cite{Martin07science,Sansonenature2010,Lafosse03jpb}. 

Here we show experimental evidence for a non-invasive and much more fundamental way to break the symmetry of ${\textrm{H}_2}^+$ avoiding any external fields and occuring in the absence of doubly excited states. We demonstrate experimentally that the transient field of the photoelectron which is ejected when an ${\textrm{H}_2}^+$ ion is created by  photoionization  is sufficient to preferentially localize the bound electron at one side of the molecule. This retroaction of the photoelectron onto its parent molecule has recently been suggested in pioneering theoretical work by Serov and Kheifets \cite{Serov14pra} but has never been recognized in an experiment \cite{Lafosse03jpb,Hikosaka03,Dowek10prl,Torres14pra}. 

In a broader context the influence of a photoelectron onto its emission source has been discussed for the photoelectric effect in solids, in particular for conducting surfaces. There it is obvious that the photoelectron will induce a positive mirror charge in the conductor. For this to happen a reservoir of highly movable conduction band electrons is necessary. The time scale on which such mirror charges are formed is still under dispute. For electrons in molecules such mobility and the ability of the bound electrons to react is hindered by the absence of bands. The essence of the effect is still captured by the concept of polarizability. An escaping electron will transiently polarize the molecular ion left behind. If the emitted electron is slow, the tail of its Coulomb potential is still significantly present while the dissociation of the molecule occurs. In such case, one might envision that such polarization can freeze out and the charge can become unequally distributed on the fragments, even for homonuclear diatomic molecules. While this may sound obvious from a general perspective, no such observation of broken symmetry has been reported so far \cite{Lafosse03jpb}. 

\begin{figure}[H]
\centering
\includegraphics[width=0.85\columnwidth]{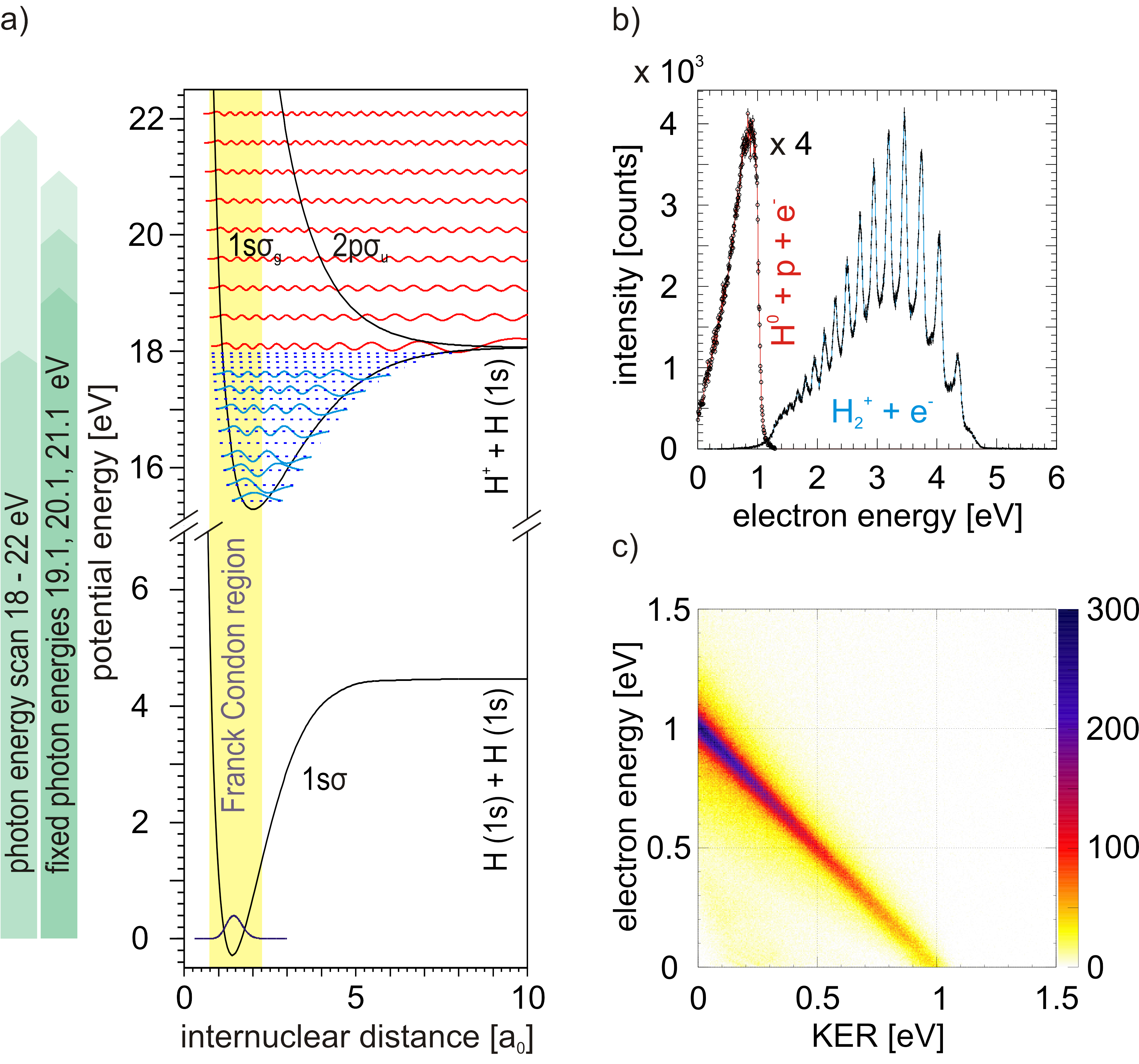}
\caption{a) Relevant potential energy surfaces for the investigated reaction channels; bound states (blue) and continuum states (red); the green arrows indicate the applied photon energies. b) Measured electron energy distribution for $E_{\gamma} = 19.1$ eV: red line corresponds to reaction channel (\ref{diss}) and is multiplied by 4, blue line shows the electron energy for the breakup according to reaction (\ref{ioni}). The ion mass enables the separation of both channels. c) Correlation between electron energy and KER in channel (\ref{diss}) for $E_{\gamma} = 19.1$ eV.}
\label{fig1}
\end{figure}

We have used photo absorption of linearly polarized photons in the range of $E_{\gamma}=19.1$ eV to $21.1$ eV to photoionize ${\textrm{H}_2}$. In this energy range two reaction channels are energetically open:
\begin{eqnarray}
\gamma + H_2 \rightarrow e^- + H_2^+(\nu) \label{ioni}\\
\gamma + H_2 \rightarrow e^- + H + p^+    \label{diss}
\end{eqnarray}  
The relevant potential energy surfaces and the measured electron energy distributions for a photon energy $E_\gamma=19.1$ eV are shown in Figure \ref{fig1}a. The dominant channel is ionization, leaving a bound but vibrationally excited ${\textrm{H}_2}^+$  behind (Eq. \ref{ioni}). The electron energy distribution reflects the distribution of vibrational states (see Fig. 1b). There is less than 5\% of Franck-Condon overlap of the ${\textrm{H}_2}$ ground state with the continuum states of ${\textrm{H}_2}^+$  at small internuclear distances \cite{Dunn66jcp}. Here, the ${\textrm{H}_2}^+$  will dissociate (Eq. \ref{diss}) and it is this small fraction of events which we will investigate further. 

Measurements were carried out at beamline UE112-PGM-1 of the synchrotron radiation scource at the Helmholtz-Zentrum Berlin
 in single bunch operation using the COLTRIMS technique \cite{doerner00pr,Ullrich03rpp}. The photon beam was crossed with a supersonic ${\textrm{H}_2}$ gas jet. The molecules in this jet are in the vibrational ground state. Electrons and ions formed in the overlap region of the photon and the molecular beam were guided by a 6 V/cm electric field onto two microchannel plate detectors  ($4\pi$ collection solid angle) with hexagonal delayline position sensitive readout \cite{Jagutzki02ieee}. 
All three components of the electron and ion momentum vectors are obtained from the times-of-flight, the positions of impact on the detector, and the ion mass. For channel (\ref{diss}) the neutral fragment is not detected. Its momentum vector can be determined from the proton and the electron momentum using momentum conservation. We have performed experiments at fixed photon energies of 19.1, 20.1, 21.1 eV and by scanning the photon energy from 18 to 22 eV. 

Figure 1c shows the correlation between the electron energy and the kinetic energy release (KER) for channel (\ref{diss}), which is the sum of the proton and H kinetic energy. The diagonal structure indicating a constant sum of all kinetic energies at $KER + E_e = E_\gamma - E_{diss}$ results from energy conservation where $E_{diss} = 18.075$ eV is the ionization potential of ${\textrm{H}_2}$ plus the dissociation energy of ${\textrm{H}_2}^+$ \cite{sharp71}. The distribution peaks at $KER = 0$ with a smooth decrease towards higher KER. 
The width of the diagonal line is mainly given by the momentum resolution of our spectrometer, which for both particles is best at zero. To make best use of this high resolution at low energy we use energy conservation and calculate $KER=(\frac{KER_m - E_{e,m}}{KER_m + E_{e,m}} + 1) \cdot \frac{E_{\gamma} - E_{diss}}{2}$. Here,  $KER_m$ corresponds to the KER calculated from the center of mass motion of the system and $E_{e,m}$ is the measured electron energy.

\begin{figure}[H]
\centering
\includegraphics[width=0.85\columnwidth]{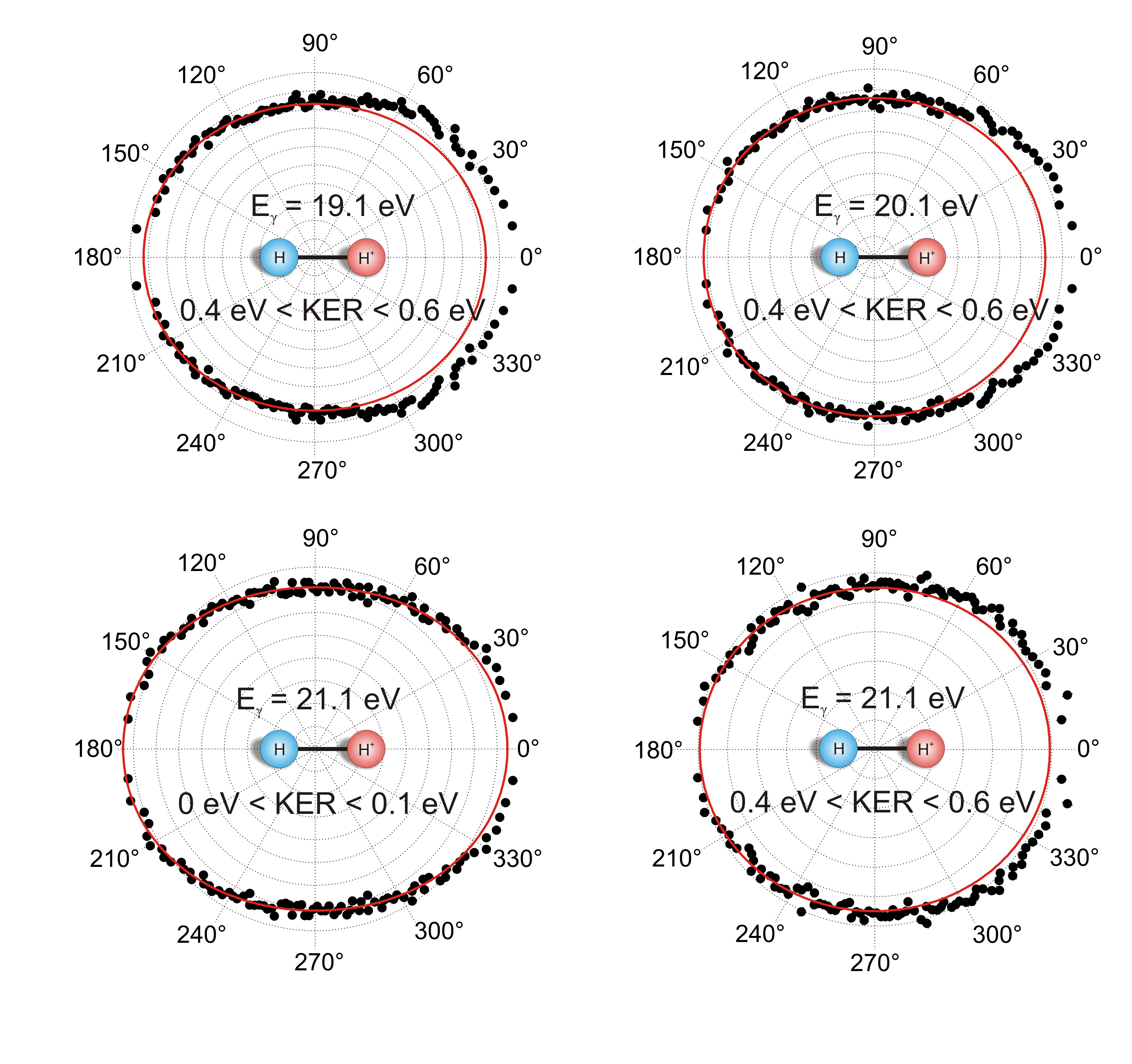}
\caption{Angular distribution of the ejected photoelectron: Shown is the angle between the electron momentum vector $\vec{k}_e$ and the molecular axis for photon energies $E_{\gamma}=19.1$, $20.1$ and $21.1$ eV. The KER is restricted to intervals from $0$ to $0.1$ eV  and from $0.4$ to $0.6$ eV respectively. The red line is a quadratic function of the form $a + b(cos(\theta))^2$ fitted to the data in the interval from $90^\circ$ to $270^\circ$ to guide the eye. The molecular orientation is fixed as shown in the middle of the picture. The statistical error bars which are not visible are smaller than the symbol size. For each histogram, the data points and the fit are mirrored at the horizontal axis for better visual inspection.}
\label{fig2}
\end{figure}

The widely used two step model of molecular photoionization assumes that the process can be split in an ionization step in which the photoelectron escapes from the molecule, leaving it in a superposition of states given by the Franck-Condon principle. In a second step the molecular ion then evolves according to its potential energy surface and the composition of the nuclear wave packet created by the preceding ionization step. The ${\textrm{H}_2}^+$ (see Fig. 1a) on the ground state $1s\sigma_g$ potential curve is the only one which can lead to low energy KER.  At photon energies close to threshold this restriction to the $1s\sigma_g$ ionic state is further  corroborated by the vanishing  Franck-Condon overlap of the ${\textrm{H}_2}$ ground state wave function with the energetically accessible part of the $2p\sigma_u$ nuclear wave function. Thus for a KER smaller than 2 eV according to the two step model the photoelectron is described by a wave function of pure ungerade parity and the ion by a wave function of pure gerade parity.
It has therefore been implicitly assumed or even concluded in several experimental \cite{Lafosse03jpb,Hikosaka03,Dowek10prl,Torres14pra} and theoretical \cite{Cherepkov03jpb,Martin09njp} studies that the electron angular distribution in the molecular frame should be symmetric with respect to the p and H side of the fragmentation. This consensus has only recently been challenged theoretically by Serov and Kheifets \cite{Serov14pra}. In order to study such possible asymmetries, we plot the angular distribution of the p-H breakup in a coordinate frame where the x-axis is given by the molecular axis (figure \ref{fig2}). The momentum of the electron $\vec{k}_e^{lab}$ is small compared to the momenta of the heavy fragments, but there is still a difference between the proton momentum with respect to the laboratory frame $\vec{k}_p^{lab}$ and the proton momentum in the center of mass of the p-H system $\vec{k}_p^{CM} = \vec{k}_p^{lab} - 0.5 \cdot \vec{k}_e^{lab}$ as noted by \cite{Hikosaka03}. We follow \cite{Hikosaka03} and plot the angle 
between the photoelectron momentum and the molecular axis, given by $\vec{k}_p^{CM}$.
We integrate over all directions of the polarization and the photon propagation. Our data show a significant asymmetry of the p-H breakup for very slow photoelectrons. The asymmetry decreases with increasing electron energy. Note that energy conservation couples the electron energy and the KER, as these data are taken at a fixed photon energy (see Figure \ref{fig1}c) 

Our data show a clear preference for the bound electron to localize during the dissociation at the proton opposite to the direction of the photoelectron. We emphasize that our setup has a collection solid angle of $4\pi$ for electrons and ions. Therefore, we can cross-check our data for any possible instrumental source of asymmetry. We have confirmed that the asymmetry flips sides
when we select electron emission to the left/right or up/down in the laboratory frame (not shown). 

To elucidate the origin of the observed symmetry breaking we study its dependence on KER and the electron energy. For a more qualitative assessment of its strength, we define an asymmetry parameter 
\begin{eqnarray} \delta=\frac{n_p - n_H}{n_p + n_H} \label{delta} 
\end{eqnarray} 
where $n_p$ and $n_H$ are the count rate for break of the p-H bond with the proton towards and opposite to the electron respectively. Thus, $\delta>0$ corresponds to the case of the proton emerging in the same hemisphere as the electron (see Fig. 2) and corresponds to $\beta=\tilde{\beta}/{E_e}$ in [10]. For channel (\ref{diss}), we show the asymmetry parameter as function of KER at three photon energies: 1, 2 and 3 eV above threshold. As KER and electron energy are related by energy conservation, the electron energy corresponding to each photon energy is plotted on an additional axis. The amount of asymmetry rises consistently with decreasing electron energy and increasing KER. The full lines show the prediction from \cite{Serov14pra}. The validity range of this calculation which treats the electron classically is restricted to $E_e >> KER$. We have therefore cut down the lines showing the theory at $E_e= \frac{1}{3} KER$. The general trend of the data and the sign and overall size of the effect is well predicted by the very approximate calculation in [10].
In this figure, both KER and $E_e$ vary as the photon energy is fixed. To unravel if the change of the asymmetry is caused by the electron energy as expected for a polarization effect and from \cite{Serov14pra}, we  have performed an additional experiment in which we scanned the photon energy. This allows to plot the asymmetry as a function of electron energy for a fixed value of the KER in Figure \ref{fig4}. 
The modeling of a retrocation of the photoelectron onto the dissociation in [10]  predicts that the asymmetry is inversely proportional to the electron energy for all KERs. Our data nicely confirm that prediction as shown by the hyperbolic fit to our data in this figure.

As argued above ionization and dissociation in two independent steps would lead to symmetric angular distributions. Thus the validity of the two step model for $\text{H}_2$ at threshold is clearly disproven by our data. We suggest that the observed symmetry breaking is induced by a retroaction of the photoelectron onto the dissociating ${\text{H}_2}^+$. This is supported by the qualitative agreement of our data with the predictions in from [10]. There the effect of the retroaction is calculated in the simple approximation of a classical electron creating a time depend field which acts on the molecular wave packet as it dissociates on the potential energy surfaces shown in Fig. 1a. This model assumes that initial conditions of the wave packet are given by the Franck-Condon overlap with the $\text{H}_2$ ground state. The wave packet evolves initially on the bound and the continuum states of $1s\sigma_g$. The electron creates a time dependent field which decreases as the electron moves away. Once the nuclear wave packet has moved out to a region where the $2p\sigma_u$ potential energy curve approaches that of the ground state the field of the electron couples bound vibrational state and the continuum states of  $1s\sigma_g$ to the $2p\sigma_u$ continuum. Therefore the nuclear wave packet is in a superposition of gerade and ungerade states which describes the localization of the bound electrons. In this scenario the amount of coupling to the ungerade state and thus the asymmetry will increase with KER. This is because the high energy part of the wave packet reaches the distance at which the coupling to the $2p\sigma_u$ occurs earlier when the electron is still closer. This trend of an increase of $\delta$ with KER is confirmed by the data in Fig. 3. The second scaling one can expect from this model is a decrease of the asymmetry with increasing electron energy. According to \cite{Serov14pra} this decrease is inversely proportional to the electron energy and again in agreement with our observation in Fig. 4. 

If one would describe the electron quantum mechanically one can expect a time dependence which mirrors that of the ionic bound part. Initially the bound electron is part of the entangled two-electron wave function of $\text{H}_2$. At all times the parity of the two-electron wave function after photon absorption is ungerade. After some time the two-electron wave function will factorize in a bound gerade and a continuum part at larger distances which is ungerade. If one would perform a measurement at this time one would find $\delta=0$. When the nuclei have separated to the region where the  $1s\sigma_g$ and $2p\sigma_u$ come close, the electron-electron interaction will entangle the wave function of the bound and the free electron. The two-electron wave function is that of a Bell state of total ungerade symmetry \cite{Schoeffler08science}. This entanglement will survive the dissociation and will lead to the measured angular correlations.

In conclusion we have demonstrated the retroaction of an escaping photoelectron onto its source. In molecules which dissociate after single photoionization, the effect leads to a preferential localization of the remaining bound electron on a site opposite to the continuum electron. For very low energetic photoelectrons, the escaping electron and the fragmentation of the molecule cannot be treated separately and the process can no longer be classified as a Franck-Condon transition. While we have observed this effect in ${\textrm{H}_2}$, the simplest system where it can occur, we speculate that the effect is general for all symmetric molecules and for all processes ejecting an electron. 

We expect that ionization by a strong laser field or by electron or ion impact as well as the dissociative ionization of heavier molecules will show similar effects if the escaping electron is slow enough. For larger molecules than diatomics we expect that the retroaction effect shown here will also influence which of several energetically degenerate bonds will break. An example would be the question which proton is ejected in a deprotonation of symmetrical hydrocarbons, which might be determined by a slow escaping photoelectron inducing a polarization in the molecular ion.

\begin{figure}[H]
\centering
\includegraphics[width=0.85\columnwidth]{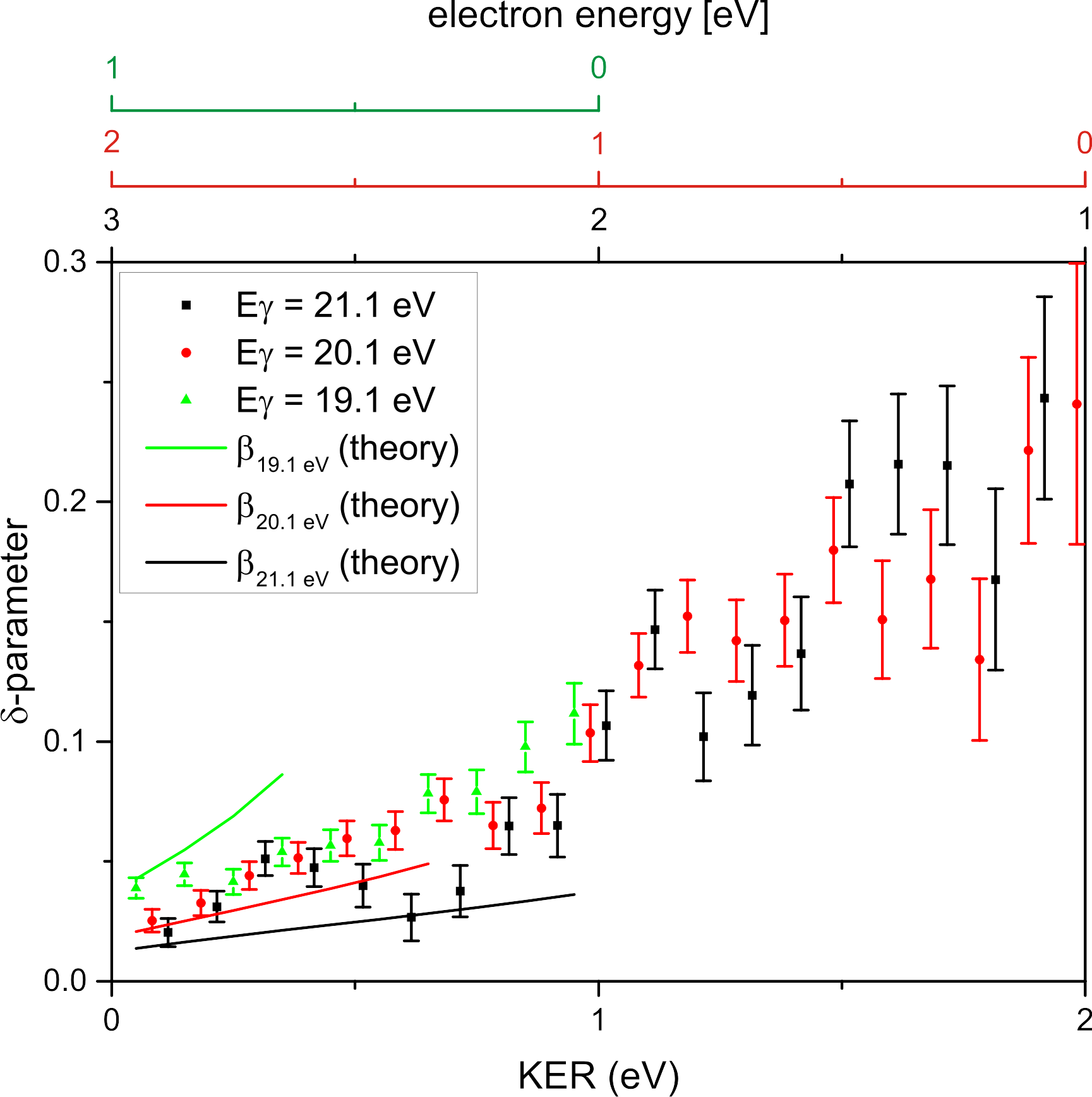}
\caption{The asymmetry parameter $\delta$ (see Equation (\ref{delta})) as a function of kinetic energy release KER for three different photon energies. Experiment: $E_{\gamma} = 19.1$ eV (green triangles), $20.1$ eV (red circles) and $21.1$ eV (black squares). Lines: corresponding predictions from \cite{Serov14pra}, colour coding as for the experiment. The theoretical curves are shown for $E_e > 2 * KER$ as the validity range of the calculation is $E_e >> KER$.}
\label{fig3}
\end{figure}

\begin{figure}[H]
\centering
\includegraphics[width=0.85\columnwidth]{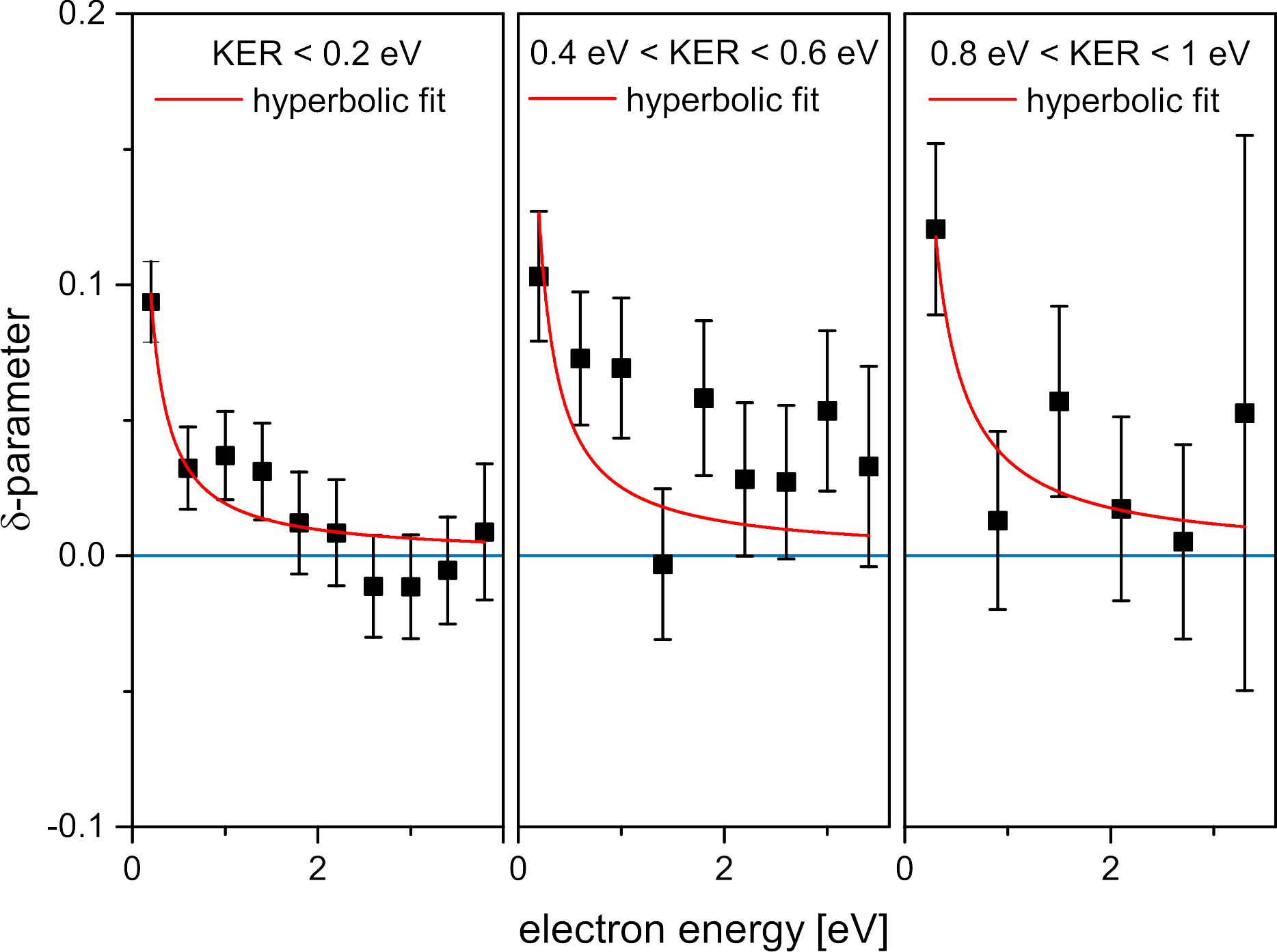}
\caption{Asymmetry parameter $\delta$ (see Equation (\ref{delta})) as a function of electron energy for $KER < 0.2 \, \text{eV}$ (left panel), $0.4 \, \text{eV} < KER < 0.6 \, \text{eV}$ (center) and $0.8 \, \text{eV} < KER < 1 \, \text{eV}$ (right panel). The asymmetry decreases with higher electron energies, and the trend is in good agreement with the hyperbolic decline $\delta \propto \frac{const.}{E_e}$ predicted by \cite{Serov14pra} (red curve).}
\label{fig4}
\end{figure}	

\acknowledgments We acknowledge helpful discussions with Anatoli Kheifets, Vladislav Serov, Ricardo Diez Muino, Fernando Martin, Sergey Semenov, Tom Rescigno and Bill McCurdy. This work was funded by the Deutsche Forschungsgemeinschaft. 
We are grateful to the staff of BESSY for excellent support during the beamtime. 
F. T. and M. W. thankfully acknowledge the financial support by HZB. We thank HZB for the allocation of synchrotron radiation beamtime.

\bibliographystyle{unsrt}
\bibliography{my_references}

\begin{thebibliography}{21}

\bibitem{Kling06science}
M.~F. Kling, C.~Siedschlag, A.~J. Verhoef, J.~I. Khan, M.~Schultze, T.~Uphues,
  Y.~N, M.~Uiberacker, M.~Drescher, F.~Krausz, and M.~Vrakking.
\newblock {\em Science}, \textbf{312}:246, 2006.

\bibitem{Ray09prl}
D.~Ray, F.~He, S.~De, W.~Cao, H.~Mashiko, P.~Ranitovic, K.~P. Singh,
  I.~Znakovskaya, U.~Thumm, G.~G. Paulus, M.~F. Kling, I.~V. Litvinyuk, and
  C.~L. Cocke.
\newblock {\em Phys. Rev. Lett.}, \textbf{103}:223201, 2009.

\bibitem{Wu13pra}
J.~Wu, A.~Vredenborg, L.~Ph.~H. Schmidt, T.~Jahnke, A.~Czasch, and R.~D\"orner.
\newblock {\em Phys. Rev. A}, \textbf{87}:023406, 2013.

\bibitem{Sansonenature2010}
G.~Sansone, Kelkensberg F., J.~F. Perez-Torres, F.~Morales, M.~F. Kling,
  W.~Siu, O.~Ghafur, P.~Johnsson, M.~Swoboda, E.~Benedetti, F.~Ferrari,
  F.~Lepine, J.~L. Sanz-Vicario, S.~Zherebtsov, I.~Znakovskaya, A.~L'Huillier,
  M.~Y. Ivanov, M.~Nisoli, F.~Martin, and M.~J.~J. Vrakking.
\newblock {\em Nature}, \textbf{465}:763--766, 2010.

\bibitem{Singh10prl}
K.~P. Singh, F.~He, P.~Ranitovic, W.~Cao, S.~De, D.~Ray, S.~Chen, U.~Thumm,
  A.~Becke, M.~Murnane, H.~Kapteyn, I.~V. Litvinyuk, and C.~L. Cocke.
\newblock {\em Phys. Rev. Lett.}, \textbf{104}:023001, 2010.

\bibitem{Fischer10prl}
B.~Fischer, M.~Kremer, Th. Pfeifer, B.~Feuerstein, V.~Sharma, U.~Thumm, C.~D.
  Schr\"oter, R.~Moshammer, and J.~Ullrich.
\newblock {\em Phys. Rev. Lett.}, \textbf{105}:223001, 2010.

\bibitem{Wu13natcomm}
J.~Wu, M.~Magrakvelidz, L.~P.~H. Schmidt, M.~Kunitski, T.~Pfeifer,
  M.~Sch\"offler, M.~Pitzer, M.~Richter, S.~Voss, H.~Sann, H.~Kim, J.~Lower,
  T.~Jahnke, A.~Czasch, U.~Thumm, and R.~D\"orner.
\newblock {\em Nature Communications}, \textbf{4}:2177, 2012.

\bibitem{Martin07science}
F.~Martin, J.~Fernandez, T.~Havermeier, L.~Foucar, Th. Weber, K.~Kreidi,
  M.~Sch\"offler, L.~Schmidt, T.~Jahnke, O.~Jagutzki, A.~Czasch, E.~P. Benis,
  T.~Osipov, A.~L. Landers, A.~Belkacem, M.~H. Prior, H.~Schmidt-B\"ocking,
  C.~L. Cocke, and R.~D\"orner.
\newblock {\em Science}, \textbf{315}:629--633, 2007.

\bibitem{Lafosse03jpb}
A.~Lafosse, M.~Lebech, J.~C. Brenot, P.~M. Guyon, L.~Spielberger, O.~Jagutzki,
  J.~C. Houver, and D.~Dowek.
\newblock \textbf{36}:4683, 2003.

\bibitem{Serov14pra}
V.~Serov and A.~S. Kheifets.
\newblock {\em Phys. Rev. A}, \textbf{89}:031402, 2014.

\bibitem{Hikosaka03}
Y.~Hikosaka and J.~H.~D. Eland.
\newblock {\em J.Elec.Spec.and.Rel.Phen.}, \textbf{133}:77, 2003.

\bibitem{Dowek10prl}
D.~Dowek, J.~F. P\'erez-Torres, Y.~J. Picard, P.~Billaud, C.~Elkharrat,
  J.~Houver, J.~L. Sanz-Vicario, and F.~Martin.
\newblock {\em Phys. Rev. Lett.}, \textbf{104}:233003, 2010.

\bibitem{Torres14pra}
J.~F. P\'erez-Torres, J.~L. Sanz-Vicario, K.~Veyrinas, P.~Billaud, Y.~J.
  Picard, C.~Elkharrat, S.~Poullain, N.~Saquet, L.~Lebech, J.~Houver,
  F.~Martin, and D.~Dowek.
\newblock {\em Phys. Rev. A}, \textbf{90}:043417, 2014.

\bibitem{Dunn66jcp}
G.~H. Dunn.
\newblock {\em J. Chem. Phys.}, \textbf{44}:2592--2594, 1966.

\bibitem{doerner00pr}
R.~D\"orner, V.~Mergel, O.~Jagutzki, L.~Spielberger, J.~Ullrich, R.~Moshammer,
  and H.~Schmidt-B\"ocking.
\newblock {\em Physics Reports}, \textbf{330}:96--192, 2000.

\bibitem{Ullrich03rpp}
J.~Ullrich, R.~Moshammer, A.~Dorn, R.~D\"orner, L.~Ph. Schmidt, and
  H.~Schmidt-B\"ocking.
\newblock {\em Rep. Prog. Phys.}, \textbf{66}:1463--1545, 2003.

\bibitem{Jagutzki02ieee}
O.~Jagutzki, A.~Cerezo, A.~Czasch, R.~D\"orner, M.~Hattass, M.~Huang,
  V.~Mergel, U.~Spillmann, K.~Ullmann-Pfleger, Th. Weber, H.~Schmidt-B\"ocking,
  and G.~D.~W. Smith.
\newblock {\em IEEE Transact. on Nucl. Science}, \textbf{49}:2477, 2002.

\bibitem{sharp71}
T.~E. Sharp.
\newblock {\em Atomic Data}, \textbf{2}:119--169, 1971.

\bibitem{Cherepkov03jpb}
S.~K. Sememov and N.~A. Cherepkov.
\newblock {\em J. Phys. B}, \textbf{36}:1409--1422, 2003.

\bibitem{Martin09njp}
J.~Fern\'{a}ndez and F.~Mart\'{i}n.
\newblock {\em New J. Phys.}, \textbf{11}:043020, 2009.

\bibitem{Schoeffler08science}
M.~S. Schoeffler, J.~Titze, N.~Petridis, T.~Jahnke, K.~Cole, L.~Ph.~H. Schmidt,
  A.~Czasch, D.~Akoury, O.~Jagutzki, J.~B. Williams, N.~A. Cherepkov, S.~K.
  Semenov, C.~W. McCurdy, T.~N. Rescigno, C.~L. Cocke, T.~Osipov, S.~Lee, M.~H.
  Prior, A.~Belkacem, A.~L. Landers, H.~Schmidt-Boecking, Th. Weber, and
  R.~Doerner.
\newblock {\em Science}, \textbf{320}:920, 2008.

\end{thebibliography}

\end{document}